\documentclass[doublecol,graphicx,times]{epl2} 
\usepackage{graphicx}
\usepackage{amsmath}
\usepackage{amsfonts}
\usepackage{amssymb}
\usepackage{amsbsy}

\def\apj{{Ast\-ro\-phys.\ J.}}

\def\grl{{Geo\-phys.\ Res.\ Lett.}}

\def\jgr{{J.\ Geo\-phys. Res.}}
\def\nat{{Nature}}
\def\prl{{Phys.\ Rev.\ Lett.}}

\def\prl{{Phys. Rev. Lett.}}

\newcommand{\thetabn}{\theta_{Bn}}

\title{A note on the theory of transverse diffusion in shock particle acceleration}
\shorttitle{Anomalous diffusion} 

\author{R. A. Treumann\thanks{Present address: International Space Science Institute, Bern, Switzerland, e-mail: treumann@issibern.ch}\inst{1,2}}
\shortauthor{R. A. Treumann}

\institute{ 
  \inst{1}Department of Geophysics, Munich University, Theresienstr. 41, D-80333 Munich, Germany\\                   
  \inst{2}Department of Physics and Astronomy, Dartmouth College, Hanover, NH 03755\\
}

\abstract{
We investigate the role of the form of the spatial diffusion coefficient in shock acceleration of fast particles. Referring to non-classical diffusion and using the results of numerical (hybrid) simulations tailored for the downstream shock population in quasi-perpendicualr high-Mach number collisionless shocks to which we apply the theory, we demonstrate that the inferred diffusion coefficients are in excellent agreement with the requirements of the theory and its predictions. Diffusion in the collisionless regime turns out to be non-classical Gibbsian (L\'evy flight), time-dependent though weak.}

\begin{document}

\maketitle
\section{\textsf{Introduction}} Diffusive shock acceleration theory of particles out of the thermal flow to high energies makes extensive use of spatial diffusion coefficients, assuming that these coefficients are generated self-consistently in the interaction of plasma waves and particles.  The multiple transitions of a fast particle back and forth across the shock under the assumption that the scattering is a stochastic process suggests a diffusive approach to the acceleration mechanism.  In the simple model where the waves propagate just parallel and antiparallel to the magnetic field this diffusion tensor is given by
\begin{equation}\label{chap6-eq-difftens}
\textsf{K}=\left(
\begin{array}{lcr}
\kappa_\| & 0  & 0  \\
0  & \kappa_{\perp,1}  & -\kappa_A  \\
0  &  \kappa_A & \kappa_{\perp,2}  
\end{array}
\right)  
\end{equation}
where $\kappa_\|\approx \frac{1}{3}\lambda v, |\kappa_A|\approx \frac{1}{3}vr_{ci}$.
These expressions hold under the assumption that $\nu_\mu/\omega_{ci}\ll 1$, with $\nu_\mu=2D_{\mu\mu}/(1-\mu^2)$ the pitch-angle scattering frequency, $\mu=\cos\alpha$, $D_{\mu\mu}$ the pitch angle diffusion coefficient, and $\alpha$ the pitch angle. Then $|\kappa_A|/\kappa_\|\ll 1$ as well. Usually the perpendicular diffusion coefficient $\kappa_\perp$ is also neglected. However, while the former can be justified somehow, $\kappa_\perp$ is implicitly determined \cite{Matthaeus2003} by an expression like
\begin{equation}\label{kappaint}
\kappa_\perp=\frac{av^2}{3B^2}\int\frac{S({\bf k}){\rm d}^3{\bf k}}{k_\perp^2\kappa_\perp+k_\|^2\kappa_\|+v/\lambda_{b\|} +\gamma({\bf k})}
\end{equation}
where $a$ is a constant of proportionality that is determined from simulations, $\lambda_{b\|}$ is the parallel correlation length of the fluctuating magnetic fields, and $S({\bf k})$ is defined through $S({\bf k},t')=S({\bf k})\Gamma({\bf k},t')$, with $\Gamma=\exp[-\gamma({\bf k})t']$, being the spatial correlation of the transverse magnetic fluctuations
\begin{equation}
\langle b_x[{\bf x}(0),0]b_x[{\bf x}(t'),t']\rangle=\int R({\bf y},t')P({\bf y}|t'){\rm d}{\bf y}
\end{equation}
$R({\bf y},t')$ is the two-point, two-time correlation function, and $P({\bf y}|t')$ is the probability  density for a particle at time $t'$ being displaced by the amount ${\bf y}$ in transverse direction. Moreover, the two-time parallel-velocity autocorrelation is modelled by the isotropic assumption $\langle v_\| (0)v_\| (t')\rangle=(v^2/3)\exp(-v_{t'}/\lambda_{b\|})$, and it is assumed that $\gamma=0$. With $P$ being a symmetric Gaussian distribution for the particle trajectory (assuming that the displacement is at all times diffusive), the average over the exponential wave factor $\langle\exp[i{\bf k\cdot x}(t')]\rangle$ which is introduced through the magnetic field fluctuations, yields just the cumulative average $\langle\exp[i{\bf k\cdot x}(t')]\rangle=\exp[(-k_\perp^2\kappa_\perp-k_\|^2\kappa_\|)t']$.

This Non-Linear Gyro-Center (NLGC) diffusion model is close to reality as it takes account not only of the random walk of the fluctuating magnetic field lines but also of the distortion of the particle orbits in this random walk. Diffusion of particles is caused by decorrelation of their orbits from the fluctuating magnetic field. A further development in this theory  \cite{Zank2004} does not use just Alfv\'enic turbulence but takes into account also the contribution of compressive modes, but the deviations from the NLGC model and the improvement over it are small and can mostly be neglected. 

The agreement between the NLGC diffusion model and the two-dimensional numerical simulation based on two-dimensional magnetic field turbulence is reasonable when compared with other models. However, the perpendicular diffusion coefficient is considerably less than quasilinear theory predicts, and the deviations from quasilinear theory become susceptible already at $r_{ci}\lesssim 0.1\lambda$. Still this model is not yet self-consisten as it ignores the feedback of the particles on the magnetic field turbulence which -- at least to some degree -- is a function of the presence of the energetic particles, as suggested by self-consistent theory \cite{Lee1982}. Determination of the diffusion coefficient at this stage is a necessary intermediate step.  A complete solution of the problem including parallel and perpendicular diffusion for the relevant waves as well as energy diffusion can be expected to come only from three-dimensional full-particle PIC simulations of shock formation including particle diffusion and acceleration to high energies. 

\section{\textsf{Non-classical diffusion (`super-diffusion')}}
 The above diffusion coefficients are all based on the assumption of classical diffusion. Since shocks are narrow transitions from one plasma state to another one particle interactions in the vicinity of shocks might be subject to statistics of extremes and not to classical statistics. In this case the diffusion becomes time-dependent as no final state is reached in the process of particle scattering. The probability $P({\bf y}|t')\to P({\bf y}|t';\nu)$ is not anymore Gaussian but has a long tail of power $-\nu$ that extends toward the rare large excursions and long waiting times between the excursions. The theory of such distributions goes back to \cite{Levy1954}. More contemporary developments based on fractal theory have been given in \cite{Shlesinger1993} and \cite{Metzler2000}, and a statistical mechanical argument has been developed in \cite{Treumann2008}. An early application of generalised Gibbsian statistical mechanics (respectively L\'evy flight theory) to the derivation of the parallel and perpendicular diffusion coefficients \cite{Treumann1997} yields a time dependent  (parallel) mean square particle displacement $\langle z(0)z(t)\rangle =\kappa_\|(d,t;\nu)t$ which leads to a time dependent parallel (or non-magnetised) particle diffusion coefficient 
\begin{eqnarray}\label{chap6-eq-alpha}
\kappa_\|(d,t;\nu)&=&\kappa_{\|{\rm cl}}[\nu/(\nu-d/2)](\nu_{\rm an} t)^\alpha \\
\alpha&=&(4\nu -2d -1)^{-1}
\end{eqnarray}
Here $d$ is the dimension of the system, $\nu_{\rm an}$ is an anomalous collision frequency that is usually nonzero, and $\kappa_{\|\rm cl}$ is the classical diffusivity based on $\nu_{\rm an}$. The condition that $P$ is a positive valued probability distribution is that $\nu-d/2>1$. The diffusion coefficient  then scales as $\kappa_\|\propto T_i\nu_{\rm an}^{\alpha-1}t^\alpha$, which for the scaling with the collision frequency gives a scaling exponent $\frac{2}{3}<|\alpha-1|<1$.  In order to find the scaling of the perpendicular diffusion coefficient one can use the ordinary classical formula 
\begin{equation}
\kappa_\perp=\kappa_\|/(1+\omega_{ci}^2/{\nu_{\rm an}}^2)
\end{equation}
(a more precise theory should use the integral equation for $\kappa_\perp$ given in Eq. (\ref{kappaint}) instead of this classical approximation) which yields 
\begin{equation}\label{chap6-eq-kappaperp}
\kappa_\perp= \frac{\nu}{\nu-d/2}\left(\frac{\nu_{\rm an}}{\omega_{ci}}\right)^{2}\,\frac{T_i}{m_i\nu_{\rm an}}(\nu_{\rm an} t)^\alpha
\end{equation}
Under the condition that $\nu_{\rm an}\propto\omega_{ci}$ this yields the scaling $\kappa_\perp\propto \kappa_{\rm B}(\omega_{ci}t)^\alpha$ with $0\leq\alpha<\frac{1}{3}$ and $\kappa_{\rm B}$ the Bohm diffusion coefficient.

 It is very important to realise that the entire physics of deviation from purely stochastic diffusion processes is contained in this surprisingly extraordinarily narrow range of exponents. Very precise determination of this exponent is therefore crucial for elucidating the actual physics of diffusion and hence also diffusive particle acceleration. Below we will discover in numerical simulations of energetic particle diffusion near shocks that the non-classical limit might indeed be realised there.
\begin{figure*}
\centerline{\includegraphics[width=0.8\textwidth,clip=]{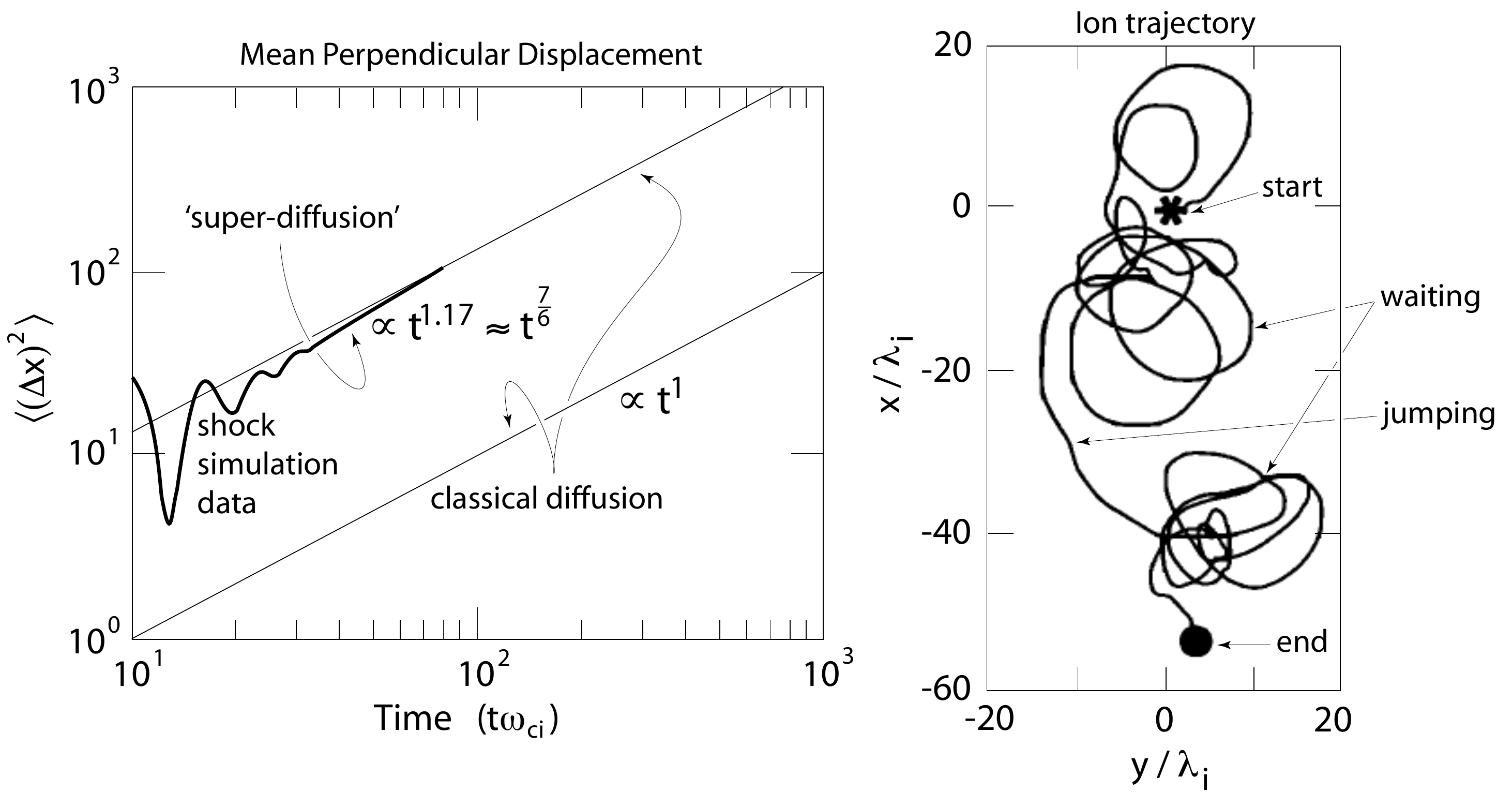}}
\caption
{\footnotesize Two dimensional numerical simulation result of the mean downstream perpendicular displacement of quasi-perpendicular ($\thetabn=87^\circ$) supercritical (${\cal M}_A=4$) shock-accelerated ions as function of simulation time (simulation data taken from \cite{Scholer2000}, courtesy American Geophysical Union). {\it Left}: The displacement performs an initial damped oscillation before settling in a continuous diffusive increase at about $t\sim 40\omega_{ci}^{-1}$. The further evolution deviates slightly from classical (linear) increase, following a $\langle\Delta x^2\rangle\propto (t\omega_{ci})^{1.17}$ power law.  This is very close to a power of $\frac{7}{6}$, suggesting that the particle diffusion process is in fact `super-diffusive' \cite{Treumann1997}. {\it Right}: Late time trajectory of an arbitrary ion  projected into the plane perpendicular to the mean magnetic field. The ion orbit in the superposition of the ambient and wave magnetic field is not a smooth stochastic trajectory. It consists of waiting (trapped gyrating) parts and parts when the ion suddenly jumps ahead a long distance as is typical for rare extreme event statistics. }\label{chap6-fig-schodiff}
\end{figure*}

This kind of transverse diffusion increases with time and, thus, the displacement of the particles grows at a faster rate than any classical transverse displacement and even faster than under Bohm diffusion $\kappa_{\rm B}$. However, if the system would have sufficient time for reaching a final state it would end up in the limit of classical diffusion with linear growing root mean square displacements. This means that under non-classical conditions the system evolves diffusively faster than it would evolve under classical conditions, while given sufficient time it would settle in classical diffusion. A diffusion of this kind is called `super-diffusion'. In an infinitely extended system it will be realised only temporarily in the initial state for times shorter than the typical classical collision time. We will return to this important statement in the discussion section. 

In a spatially limited system like the shock transition the final stationary diffusion state might not be reached, however, since the particles have not sufficient time to undergo classical collisions. In this case the above classical theories require correction for time dependent diffusion, and a stationary state is achieved only when the time dependent diffusion is balanced by losses at the moment when the mean displacement exceeds the size $\Delta_{\rm d}$ of the downstream region. From the definition of $\kappa_\perp=\Delta_{\rm d}^2/\tau_D$ one may estimate the limiting diffusion time $\tau_D$ as 
\begin{equation}
\nu_{\rm an} \tau_D =\left[\frac{\Delta_{\rm d}^2}{r_{ci}^2}\left(1-\frac{d}{2\nu}\right)\right]^\frac{1}{\alpha+1}
\end{equation}
which yields the following limitations on the anomalous collision frequency:
\begin{equation}
\left(\frac{2}{d}\frac{\Delta^2_{\rm d}}{r_{ci}^2}\right)^\frac{3}{4}<\nu_{\rm an}\tau_D<\frac{\Delta_{\rm d}^2}{r_{ci}^2}\nonumber
\end{equation}
The left limit holds for the extreme case $\nu-d/2=1$, the right for $\nu\to\infty$ (the latter corresponding to a Gaussian probability distribution). For the Earth's bow shock with the downstream region being the magnetosheath of width $\Delta_{\rm D}\sim 2\,{\rm R_E}$ the relative diffusion times will be roughly $\nu_{\rm an}\tau_D\gtrsim 2\times 10^4$ (here $1{\rm R_E}=6436$ km is the Earth radius). Measuring the limiting diffusion time $\tau_D$ of the energetic particles provides an opportunity to determine the anomalous collision frequency $\nu_{\rm an}$ that governs the diffusive interaction.

\section{\textsf{Shock simulation of particle diffusion}}
 In order to confirm the diffusive nature of the ion acceleration process in supercritical shocks,  three-dimensional numerical hybrid simulations (with particle ions and a neutralising electron fluid) have been performed \cite{Scholer2000} measuring the root-mean-square high energy ion displacement $\langle\Delta x^2\rangle$ in the direction perpendicular to the  magnetic field in the downstream region as a function of simulation time. The simulations corresponded to the region downstream of a quasi-perpendicular supercritical shock of Mach  number ${\cal M}_A=4$ and shock-normal angle $\thetabn=87^\circ$. Due to restrictions of computing power the diffusion process could be followed only just up to simulations times roughly $t\sim90\, \omega_{ci}^{-1}$. 

The result of these simulations is reproduced here in Figure\,\ref{chap6-fig-schodiff} with simulation data taken from \cite{Scholer2000}. Here $\langle\Delta x^2\rangle$ has been plotted as a function of time $t\omega_{ci}$. Initially the displacement performs a large amplitude damped oscillation until the diffusive equilibrium is attained at about $t\omega_{ci}\sim 40$. For later times the particle displacement increases continuously. However, the exponent of the increase is found not to be unity as was expected for classical diffusion. It is rather larger, being 1.17, which is very close to $\frac{7}{6}$ identifying the diffusion process as `super-diffusion'. (Note that, because of the large number $\sim6.3\times10^6$ of macro-particles used in the simulation of which 525000 have high energies and contribute to the determination of the mean displacement, and because of the high time resolution, the statistical error of the measurement is less than the line width!) In \cite{Scholer2000} this discrepancy is well noted while, nevertheless,  it is insisted on interpreting the simulations in terms of the classical diffusion picture. Here we use these data in order to check whether they fit the non-classical diffusion theory.

Above it has been shown that the exponent of the mean displacement in anomalous or `super-diffusion' is  $1+\alpha$ with $\alpha=(4\nu-2d-1)^{-1}$, as given in Eq.\,(\ref{chap6-eq-alpha}) and $0\leq\alpha\leq\frac{1}{3}$. The lower bound corresponds to classical diffusion (the linear increase of the displacement in the log-log representation of Figure\,\ref{chap6-fig-schodiff}). In the simulated case $\alpha\approx \frac{1}{6}$ lies clearly in the permitted range of exponents. We may use this value together with the known dimensionality $d=3$ of the simulation in determining the value of the anomalous parameter $\nu=\frac{13}{4}=3.25$. (If using the exact value $\alpha=0.17^{-1}$, we obtain $\nu=3.22$.) This value is sufficiently far above the marginal value of $\nu=\frac{3}{2}$ for the three-dimensional case (or $\nu=1$ for the two-dimensional case). For the basic theory see \cite{Treumann2008}. It is, however, also far enough below the classical diffusive limit $\nu\to\infty$ thus identifying the diffusion process as indeed being anomalous, non-stochastic (non-Markovian)  and super-diffusive. Since these simulations are completely collision-free, particle diffusion is entirely determined by anomalous processes that are mediated by the self-consistently excited wave spectrum downstream of the shock. 
\begin{figure*}
\centerline{\includegraphics[width=0.9\textwidth,clip=]{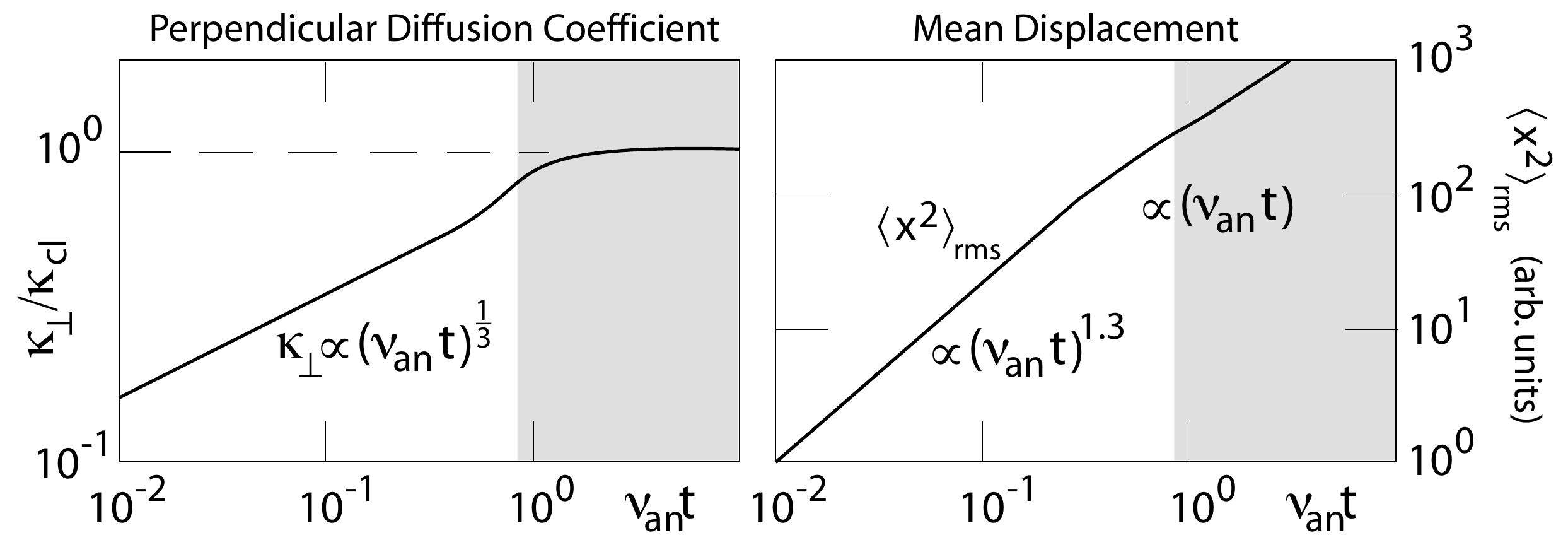}}
\caption
{\footnotesize Evolution of the anomalous perpendicular diffusion coefficient under conditions of L\'evy statistics. {\it Left}: The diffusion coefficient is a power law increasing with time $t$ with anomalous collision frequency $\nu_{\rm an}$. Its maximum slope is $+\frac{1}{3}$. Approaching collision time $\nu_{\rm an}^{-1}$ it merges into the constant classical diffusion coefficient (shaded region). {it Right}: Evolution of the rms perpendicular displacement $\langle x^2\rangle$ as function of time. The maximum slope of the power law is $\sim 1.3$. When classical diffusion takes over the displacement evolves linearly with time.}\label{chap6-fig-andiff}
\end{figure*}

In view of this reasoning it is instructive to look at the trajectory an arbitrarily chosen fast ion performs in the magnetic fluctuation field. The right part of Figure\,\ref{chap6-fig-schodiff} taken from \cite{Scholer2000} shows such an orbit in the plane perpendicular to the magnetic field at a late time $100<t\omega_{ci}<160$, when no average diffusive displacement was determined anymore. Clearly, the trajectory is far from being a smooth diffusive particle orbit. It rather consists of a sequence of gyrating sections and sections when the particle breaks out of gyration. When performing the former sections, the particle is `waiting' at its guiding centre location, approximately conserving its magnetic moment and performing a slightly modulated gyration around the magnetic field which is due to weak interaction with the wave field.  In this regime it is probably interacting only very weakly with the wave fields. During the break-out sections it suddenly jumps out in order to occupy another waiting position. The break-out is most probably caused by a brief intense wave-particle interaction.(This could have but has not been check in the simulations of \cite{Scholer2000}.) Clearly, during this break-out the magnetic moment of the particle is not anymore conserved. The instantaneous location of the chosen particle is progressing only in $x$-direction. Other particles also break out into direction $y$. On a short time scale such a process cannot really be described anymore as being stochastic. 

\section{\textsf{Discussion}}
The non-stochasticity is reflected in the time dependence of the perpendicular diffusion coefficient. However, as we have noted above, this time dependence will last only as long as the time remains to be shorter than the collision time, when the diffusion coefficient will assume its constant classical value. This reasoning implies that {\it the actual diffusion in the collisionless regime is very slow}. This has been noted (with surprise, because referring to classical diffusion) from the simulations \cite{Scholer2000}. The diffusion is in fact much weaker than classical diffusion even though the increase of the displacement is substantially faster than under classical diffusive conditions. Using classical diffusion in diffusive acceleration grossly overestimates the diffusive effect. The term `super-diffusion' does not apply to the strength of diffusion but just to its time-dependence. 

The excellent agreement between the numerical simulations and the theoretical predictions and limitations on the transverse diffusion coefficient in super-diffusion raises the question whether and for how long diffusion will remain to be anomalous. Obviously the system which we have considered is completely collisionless. Also, the simulations which we have referred to and used in order to estimate the time dependence of the anomalous diffusion coefficient, have been performed under strictly collisionless conditions. (In fact, because of computational limitations they could not be extended into longer times.) 

It is reasonable to assume that any real collisionless system will ultimately see collisions between the particles if only permitted to wait sufficiently long. This will happen when $t\sim\tau_c=\nu_c^{-1}$, where $\tau_c$ is the binary collision time, and $\nu_c$ the binary collision frequency. At this time classical diffusion sets on, and $\kappa\to\kappa_{cl}=$\,const, $\langle x^2\rangle=\langle x^2(t=\tau_c)\rangle t$. The constant of proportionality herein is the average displacement the particles have reached at time $t=\tau_c$. How this displacement is reached is, however, not described by classical theory, as in classical diffusion theory the time is measured in units of the collision time $\tau_c$. Rather non-classical diffusive processes like the one proposed in \cite{Treumann1997} and used in this Letter are responsible for the particles to provide the initial condition for classical diffusion. This is graphically shown in Figure \ref{chap6-fig-andiff}, where the evolution of the mean-square displacement and the corresponding evolution of the diffusion coefficient are shown for times $t<\tau_c$. Even though the evolution of the mean-square displacement with time is faster than under classical conditions, diffusion is weak as it reaches the initial condition for classical diffusion only at collision time $\tau_c$. 

Classical diffusion is not starting from zero as usually believed, but rather from the value of displacement provided by the non-classical diffusion process. Correspondingly, the non-classical diffusion coefficient is {\it smaller} than the classical diffusion coefficient even though it evolves in time for catching up with the classical diffusion coefficient at time $t=\tau_c$. Hence, even though non-classical diffusion evolves with time, the absolute value of the non-classical diffusion coefficient is smaller than the value of the classical diffusion coefficient in the same medium after collisional conditions are reached. Super-diffusion is a fast process, but it is weaker than classical diffusion.

We finally comment on the narrow range of powers $\alpha$ inferred both from super-diffusive non-classical diffusion\newpage\noindent  theory and  simulations. Even though the range of the realised powers is so small, it contains all of the important physics. In general power laws seem to be rather insensitive to the subtleties of the underlying physical processes. In observations this fact is generally taken not as serious. However, slight deviations in power may imply completely different physics as our example of  shock accelerated diffusion has shown. Determination of powers must be done with care, and caution must be applied in their physical interpretation. 


\end{document}